\documentclass{aa}
\usepackage{natbib}
\usepackage[varg]{txfonts}
\usepackage[colorlinks=true,linkcolor=blue,citecolor=blue,filecolor=blue,urlcolor=blue]{hyperref}
\usepackage{graphicx}
\usepackage{bm}
\usepackage{CJKutf8}
\newcommand{\fg}[1]{Figure~\ref{fig:#1}}
\newcommand{\Fg}[1]{Figure~\ref{fig:#1}}
\newcommand{\eq}[1]{Equation~(\ref{eq:#1})}
\newcommand{\eqr}[1]{Equation~\ref{eq:#1}}
\newcommand{\Tb}[1]{Table~\ref{tab:#1}}
\newcommand{\Eq}[1]{Equation~(\ref{eq:#1})}
\newcommand{\eqs}[2]{Equations~(\ref{eq:#1}) and (\ref{eq:#2})}

\newcommand{\app}[1]{Appendix~\ref{app:#1}}
\newcommand{\se}[1]{Section~\ref{sec:#1}}
\newcommand{\Se}[1]{Section~\ref{sec:#1}}

\newcommand{\oss}{OSSOS\xspace}
\newcommand{\ossp}{OSSOS++\xspace}

\begin{document}

\title{From planetesimals to dwarf planets by pebble accretion}
\author{Chris W. Ormel\inst{1}, Yukun Huang (\begin{CJK*}{UTF8}{gbsn}黄宇坤\end{CJK*}) \inst{1, 2}}
\institute{Department of Astronomy, Tsinghua University, Beijing 100084, China \\
	\email{chrisormel@tsinghua.edu.cn}
	\and
	National Astronomical Observatory of Japan, 2-21-1 Osawa, Mitaka, Tokyo 181-8588, Japan
}

\date{\today}

\abstract{The size distribution of trans-Neptunian objects (TNOs) in the Kuiper Belt provides crucial insights into the formation and evolution of the outer Solar System. Recent observational surveys, including the Outer Solar System Origins Survey (\ossp), have revealed that dynamically cold and hot TNO populations exhibit similar size distributions for dimmer objects ($H_r > 5$), which are consistent with planetesimal formation by streaming instability (SI). However, the hot population contains a significantly larger number of massive bodies, including several dwarf planets. In this study, we investigate the role of pebble accretion in shaping the size distribution of hot TNOs, after their formation in the primordial disk (PB) between 20 and 30 au and before these bodies were dynamically implanted into their current orbits by a migrating Neptune. We find that pebble accretion grows the most massive bodies only, consistent with the flattening of the distribution brightwards of $H_r=5$. All results point to a correlation (degeneracy) between the pebble aerodynamic size and the intensity of the gas motions. Nevertheless, accretion from an inward-drifting stream of pebbles is unlikely, as it puts extreme demands on the mass budget of pebbles. In particular, the masses of the cold classicals are too low to trigger pebble accretion. Accretion in an environment where pebbles are entrained, as believed to be the case in rings seen with the Atacama Large Millimeter Array (ALMA), is preferable. Combining the constraints obtained from this study with ALMA imagery morphology fitting reveals a typical pebble aerodynamic size of $\tau_s \sim 10^{-2}$, a turbulent diffusivity parameter $\alpha_D\sim10^{-3}$, and a total accreted pebble mass of ${\sim}10\,m_\oplus$ in the primordial belt. Those TNOs formed through significant pebble accretion with masses exceeding ${\sim}10^{-4}\,m_\oplus$ are likely to satisfy the International Astronomical Union's ``round shape'' criterion for dwarf planets.}

\maketitle
\section{Introduction}
The trans-Neptunian space, conventionally known as the Kuiper belt, hosts numerous icy bodies which are believed to be ancient remnants of the Solar System's outer planetesimal disk. As more trans-Neptunian objects (TNOs) are discovered by various modern observational surveys, such as the Canada–France Ecliptic Plane Survey (CFEPS; \citealt{Jones.2006}), the Deep Ecliptic Survey (DES; \citealt{Adams.2014}), the Outer Solar System Origins Survey (OSSOS; \citealt{Bannister.2018}), the Dark Energy Survey (DES; \citealt{Bernardinelli.2022}), and the DECam Ecliptic Exploration Project (DEEP; \citealt{Trilling.2024}), there is a growing consensus that the Kuiper belt consists of two distinct populations: dynamically cold and dynamically hot TNOs. 

Cold TNOs, also known as cold classicals, are characterized by their low inclinations and relatively stable orbits, suggesting that they formed in their current locations and have remained dynamically cold \citep{Petit.2011}. In contrast, hot TNOs, which include the hot classicals, resonant objects, and scattering disk objects (see \citealt{Gladman.2008} for the definitions of the TNO dynamical classes), exhibit higher inclinations and eccentricities, suggesting a more violent dynamical history. Moreover, cold TNOs tend to have redder colors, indicating a different surface composition and evolution history compared to the more neutral colors of hot TNOs \citep{Doressoundiram.2002, Trujillo.2002, Pike.2017, Schwamb.2019, Muller.2020, Fernandez-Valenzuela.2021}. The binary fraction is also higher among cold TNOs \citep{Noll.2020}, especially those of comparable sizes. Additionally, dwarf planet-sized TNOs are predominantly found in the hot populations, with cold TNOs lacking large objects \citep{Brown.2008, Fraser.2014}. All these differences—orbital characteristics, colors, binary fractions, and size distributions—support the hypothesis that cold and hot TNOs formed at different radial locations and experienced different dynamical evolutions. Cold TNOs likely formed in situ at their current locations ($\sim$45~au), while hot TNOs formed closer to the Sun  ($\sim$20~au) and were subsequently implanted into their current orbits during the migration of the giant planets \citep{Malhotra.1993, Gomes.2003, Levison.2008, Nesvorny.2015}.

The implantation process of hot TNOs is closely linked to the planetesimal-driven migration of the giant planets, particularly Neptune. In this hypothesis, Neptune is believed to have formed closer to the Sun and migrated outward into a primordial planetesimal belt that spanned $\sim$20--30~au \citep{Gomes.2003, Levison.2008}, with a total estimated mass of ${\sim}20\,m_\oplus$ \citep{Nesvorny.2015}. Most of the planetesimals were scattered inward consecutively by Neptune, Uranus, and Saturn, until their orbits crossed the orbit of Jupiter. The strong gravity of Jupiter effectively ejected the vast majority of these planetesimals out of the Solar System, leaving only a small fraction to form the Jupiter trojans \citep{Morbidelli.2005, Nesvorny.2013}. Another fraction of the planetesimals were scattered outward by a migrating Neptune into the trans-Neptunian region, forming the hot or implanted population \citep{Malhotra.1993, Nesvorny.2016}.

As the number of discovered TNOs continues to grow, the size distribution\footnote{The absolute magnitude measured with the $r$ filter ($H_r$) is often used to study the size of TNOs. For the conversion between $H_r$ and TNO mass, see \se{hot} and \app{TNOmass}.} of both cold and hot TNOs is becoming increasingly well-constrained. \citet{KavelaarsEtal2021} reported the $H_r$ distribution of the cold classical belt from $H_r\simeq 5$ to 12 detected by \oss, and found an exponentially tapered power law with an index of $\alpha \approx 0.4$. This functional form is also consistent with the DEEP survey \citep{Napier.2024}. Recently, streaming instability (SI) has emerged as one of the leading theories for planetesimal formation in the solar nebula, as evidenced by the high binary fractions observed in cold classicals \citep{NesvornyEtal2010, NesvornyEtal2019}, and the discoveries of contact binaries such as Arrokoth \citep{McKinnonEtal2020,LyraEtal2021}. The exponentially tapered power law is also consistent with SI simulations that model the planetesimal formation process \citep{Schafer.2017, Abod.2019}.

In a follow-up study of the \ossp survey, \citet{PetitEtal2023} conducted a similar analysis for the hot belt. They found that the shape of the $H_r$ distribution of the hot belt is similar to that of the cold belt, ranging from $H_r \approx 5$ to $H_r \approx 8.3$ (corresponding roughly to bodies with $D\approx 90\,\mathrm{km}$ in diameter\footnote{The relation between the $H_r$ magnitude, mass, and diameter is detailed in \app{TNOmass}. When quoting diameters, a nominal internal density of $1\,\mathrm{g\,cm^{-3}}$ is assumed.}). This upper limit corresponds to the sensitivity threshold of the \ossp survey at $\sim$48~au, given an apparent magnitude limit of $m_r\approx25$. They also found that all the dynamically hot TNOs and those limited to the main belt
exhibit remarkably similar $H_r$ distributions (see Figure 3 in \citealt{PetitEtal2023}), further supporting the shared origin of the dynamically hot TNOs. Given the similarity in the size distributions of the cold and hot TNOs at the low-mass end ($H_r \gtrsim 5$, roughly $D \gtrsim 400$~km), it is natural to postulate that the two populations started off with similar initial mass functions (IMF) given by the SI process. In contrast, the (future) hot population had the high-mass end altered by other planet formation processes.

\begin{figure}
	\centering
	\includegraphics[width=\columnwidth]{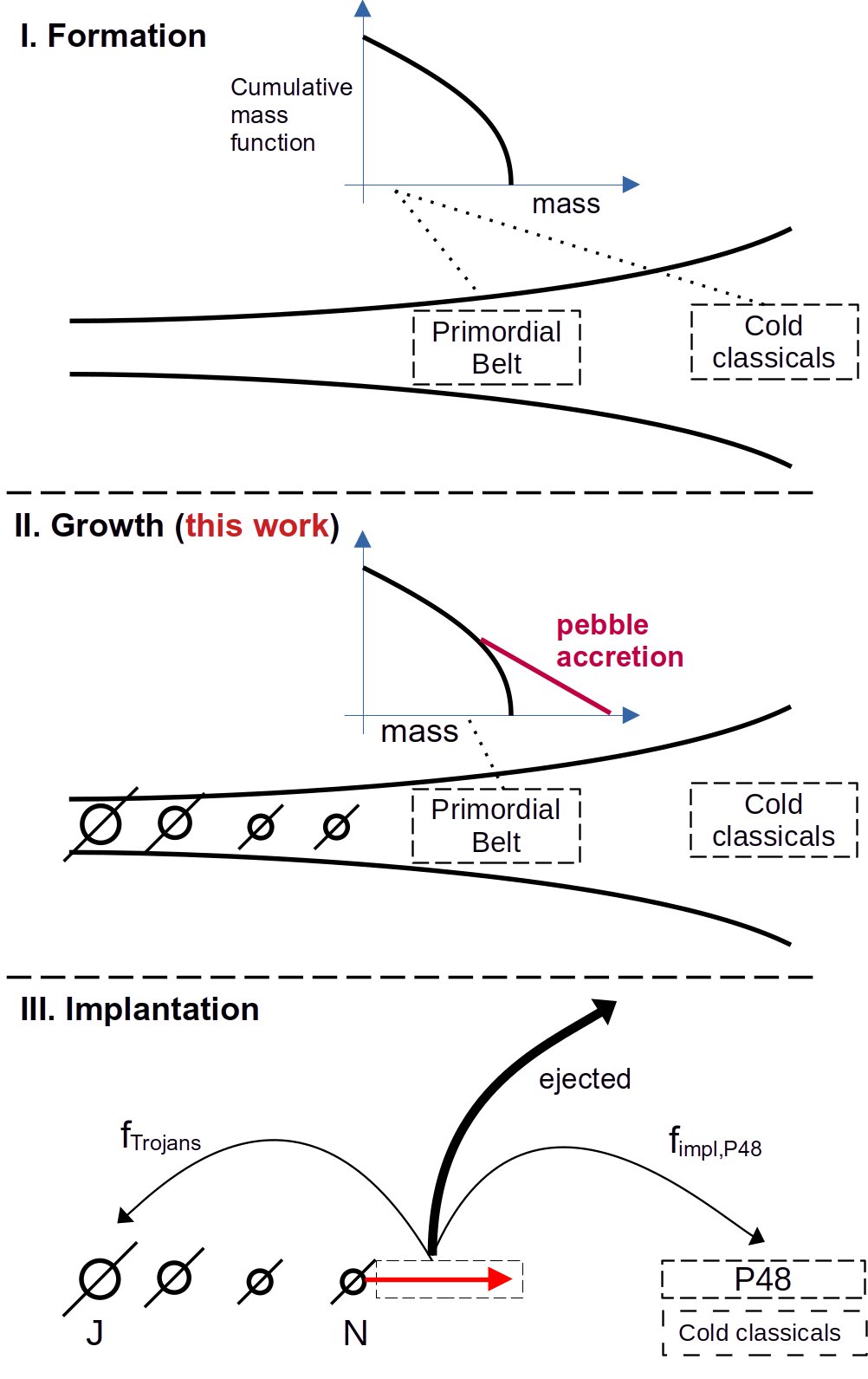}
	\caption{\label{fig:sketch}Envisioned chronology of the formation of trans-Neptunian objects. \underline{I. Formation.} An initial burst of planetesimal formation populates bodies in the primordial belt (PB) at ${\sim}$20 au and the cold classicals at ${\sim}$40--50~au). It is assumed that the shape of the cold classicals and the primordial belt size distributions are similar, both featuring a tapered exponential at high mass. \underline{II. Growth}: Pebble accretion, acting preferentially on the massive bodies, alters the size distribution of the PB. Growth takes place in a gas-rich medium. This is the phase investigated in this work. \underline{III. Implantation.} Neptune's migration disperses the PB with a implantation fraction $f_\mathrm{impl,P48}$ populating the dynamically hot bodies within 48 au (population P48).}
\end{figure}
\begin{figure*}
	\sidecaption
	\includegraphics[width=0.7\textwidth]{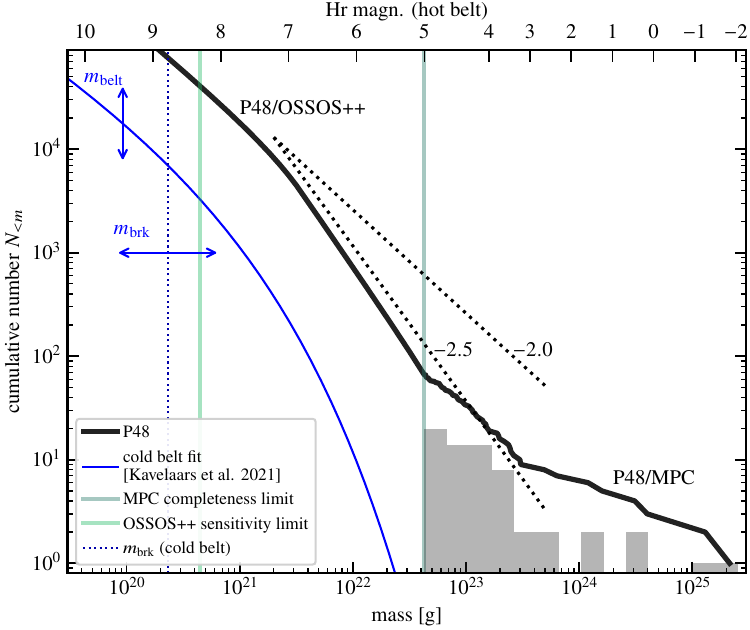}
	\caption{\label{fig:cumulative}Cumulative number distribution ($N_{<m}$) of bodies in the cold belt and the dynamical hot P48. The cold belt is well described by the analytical fit of \citet{KavelaarsEtal2021} (\eq{Kavelaars2021}), whose shape is assumed to be identical to the distribution of the bodies in the primordial belt. Population P48 is reconstructed by taking all bodies brighter than $H_r=5$ and within 48 au from the MPC database (histograms) while bodies $5 \le H_r \le 8.3$ are assumed to follow the completeness-corrected distribution from \ossp. The corresponding cumulative mass function is given by the black curve. The conversion between mass and magnitude (top) assumes an albedo typical for hot belt bodies (see \app{TNOmass}). The dotted lines labeled $p=-2.0$ and $p=-2.5$ correspond to $N_{<m} \propto m^{1+p}$ with $p=-2$ indicating an equal amount of mass per logarithmically spaced mass bin (the corresponding indices in terms of a radius spectrum $dN/dR \propto R^q$ are $q=-4$ and $q=-5.5$, respectively).
		Pebble accretion operates to selectively grow the largest TNOs, flattening the size distribution.
	}
\end{figure*}

Here we postulate that the process that affected the high-mass end of the size distribution of bodies in the primordial belt was pebble accretion. A similar suggestion has recently been made by \citet{CanasEtal2024}. Pebble accretion is a process that takes place in a gas-rich disks and involves small particles that are aerodynamicly coupled to the gas \citep{OrmelKlahr2010,LambrechtsJohansen2012}. Pebble accretion has been widely used to model the growth of planetary bodies in our Solar System and in other exoplanet systems \citep{LiuJi2020,DrazkowskaEtal2023}. Invoking pebble accretion to explain the flattening of the TNO size distribution at high mass is attractive, as pebble accretion operates on bodies above a certain mass scale \citep{Ormel2017}.

The process that is envisioned is schematically illustrated in \fg{sketch}. SI forms bodies that follow a tapered-exponential in both the cold and the primordial belts, but only the primordial belt featured conditions for pebble accretion to materialize. While the size distribution of smaller bodies stayed intact and continued to follow the initial mass function (IMF) bodies obtained after SI, the larger bodies consumed pebbles. Afterwards, the dynamical instability implanted these bodies into the trans-Neptunian space.

Pebble accretion is sensitive to the physical properties of the particles, and to the local disk conditions. Two parameters in particular are responsible for the characteristics of the accretion: the particle aerodynamic size $\tau_s$ and the velocity $\Delta v$ at which pebbles approach the planetary body. For this work, we conducted a quantitative study to verify which combination of these and other parameters are consistent with the size distribution of current TNOs. To this end, we carefully selected a sample of TNOs that are complete, based on the \ossp survey and the Minor Planet Center (MPC) database. For simplicity, we limited the number of parameters in our model (e.g., we only treated one particle size). As one of our key results, we find a tight relation between pebble aerodynamic size $\tau_s$ and relative velocity $\Delta v$, which favors formation in an environments that resemble the azimuthally symmetric structures seen in images taken by the Atacama Large Millimeter Array (ALMA; \citealt{BaeEtal2023}), commonly referred to as ALMA rings.

The structure of the paper is as follows. In \se{model} we describe our simplified model to follow the growth of planetary bodies by pebbles. The simplified nature allows exploration by Markov chain Monte Carlo (MCMC) methods (\se{results}). In \se{scenarios}, the implications of these findings are reported in the context of formation models for our Solar System. Further assessment of our model is given in \se{discussion} and the conclusions are given in \se{conclude}.

\section{Model}
\label{sec:model}
This section consists of three parts. In \se{distribution}, we reconstruct a population of TNOs that is complete, against which the model outcome can be evaluated. \Se{rates} describes the pebble accretion model, which is formulated in dimensionless units. While this approach makes the model more abstract, it avoids the need to introduce a specific disk model. Finally, in \se{parameters}, we construct the MCMC likelihood function and outline the model parameters.

\subsection{The mass function of the dynamical hot population}\label{sec:hot}
\label{sec:distribution}

In \fg{cumulative}, we have reconstructed the present-day mass distribution of dynamically hot TNOs within 48 au, a population we refer to as P48. This distribution is composed of two parts. At high masses, we take the bodies directly from the Minor Planet Center (MPC) database\footnote{The file \texttt{distant\_extended.dat} was retrieved from \url{https://www.minorplanetcenter.net/data} on November 9 2024.} and filter out the cold classicals by applying an $i_\text{free} < 4^\circ$ cut \citep{Huang.202269}. Specifically, we assume that the data, which includes the entire hot belt, inner belt, 3:2 population (plutinos), 2:1 population (twotinos), as well as scattering TNOs, is complete for $H_r < 5$ bodies within $a \le 48\,\mathrm{au}$ \citep{Weryk.2016, PetitEtal2023}.

We excluded 13 members that belong to the Haumea collisional family, as they are likely fragments ejected from Haumea during or after Neptune's migration \citep{Proudfoot.2019}, and are therefore unrelated to the prior pebble accretion process. To account for the lost mass of the Haumea impactor (which could be a moon of Haumea or a dwarf planet from the scattering disk depending on the hypothesis), we include a new body with a mass equal to 3\% of Haumea's \citep{Vilenius.2018, Pike.2020} in our sample. Additionally, we add Neptune's largest moon, Triton,\footnote{This assumes that the implantation probability of bodies ending up as TNOs is similar to the capture probability by Neptune.} which is believed to be a captured moon from the primordial belt \citep{Agnor.2006}, as well as Pluto's largest moon, Charon, which likely formed from a giant impact \citep{Canup.2005}. In total, 66 bodies constitute
the high-mass portion of the P48 population (see \Tb{tnos-mpc}), reflecting our current best understanding of the dwarf planet size distribution.

Trans-Neptunian Objects dimmer than $H_r=5$ in P48 are modeled on the debiased distribution of the hot belt\footnote{Haumea family members with $H_r > 5$ are not corrected, as the OSSOS survey reveals a significant scarcity of small members within the Haumea family \citep{Pike.2020}.}, based on the \ossp survey \citep{PetitEtal2023}. Their Figure 1 presents the differential distribution, which we fit by a broken power law
\begin{equation}
	\Delta \frac{d\log_{10}N}{dH} = \left\{ \begin{array}{ll}
		0.95H_r -3.45 & (5.0\le H_r \le6.8)  \\ \displaystyle
		0.55H_r -0.73 & (6.8\le H_r \le 8.3) \\
	\end{array} \right.
	\label{eq:dNdHfit}
\end{equation}
where $\Delta=0.25$ dex is the bin spacing adopted by \citet{PetitEtal2023}. We further estimate that the hot classicals makes up 50\% of the dynamical hot bodies in P48, based on population estimates of $H_r < 8.7$ objects: the hot belt contains $20 \times 10^3$ objects \citep{PetitEtal2023}, while the inner belt, twotinos, plutinos, and scattering populations within $a < 48$~au contain $3, 4.4, 8,$ and $3 \times 10^3$ objects, respectively \citep{Petit.2011, Volk.2016, Chen.2019}. The fraction of hot belt TNOs in P48 is thus $20/(20+3+4.4+8+3)\approx0.5$. Therefore, we multiply the number $N$ that follows from \eq{dNdHfit} by a factor 2 when reconstructing the P48 distribution (see \fg{P48}).

\begin{figure}
	\centering
	\includegraphics[width=0.9\columnwidth]{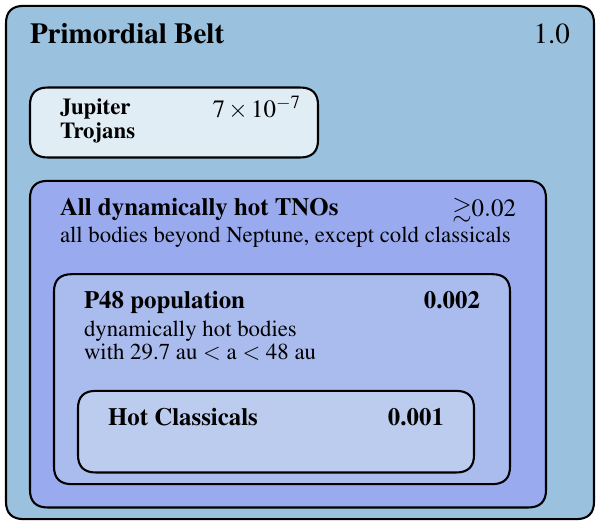}
	\caption{\label{fig:P48}Implantation fractions for bodies of the primordial belt into the trojans, hot classicals, and P48. Numbers in bold are used in this work (see text for motivation) and others are given as context. The numbers for the Jupiter trojans follow \citet{Nesvorny.2013} and those for the entire dynamically hot TNO population \citet{Nesvorny.2016} and \citet{Huang.2023t}. 
    }
\end{figure}

In \fg{cumulative}, the cumulative distribution of bodies in P48 is plotted alongside the debiased distribution of the cold belt, which is well described by an exponentially tapered power law \citep{KavelaarsEtal2021}. Converted in terms of mass (see \app{TNOmass}), the mass function reads:
\begin{equation}
	\label{eq:Kavelaars2021}
	N_{>m} = N_\mathrm{pre}
	\left( \frac{m}{m_\mathrm{brk}} \right)^{-\alpha} \exp \left[ - \left( \frac{m}{m_\mathrm{brk}} \right)^\gamma \right],
\end{equation}
where for the cold belt $\alpha=0.67$, $\gamma=0.42$, and the mass scale indicating the transition to the exponential cut-off regime, $m_\mathrm{brk} = 2\times10^{20}\,\mathrm{g}$ ($D\approx70\,\mathrm{km}$). In \eq{Kavelaars2021}, $N_\mathrm{pre}$ is a normalization factor, which, for the specified values of $\alpha$ and $\gamma$, can be expressed in terms of the total mass in the belt, $m_\mathrm{belt}$, as $N_\mathrm{pre}=m_\mathrm{belt}/2.81 m_\mathrm{brk}$.
It is important to note that converting between magnitude $H_r$ and mass depends on albedo, which varies between bodies in the hot and cold belts. Consequently, any
similarity in the magnitude distributions of the cold and hot belt populations \citep{PetitEtal2023} does not necessarily imply similarity in the mass distributions. To account for this, we treat $m_\mathrm{brk}$ as a model parameter. Along with the total initial mass $m_\mathrm{imf}$ of the primordial belt, this parameter defines the initial distribution prior to pebble accretion.

\subsection{Growth of bodies by pebble accretion}
\label{sec:rates}
Not accounting for gravitational effects, a spherical body of radius $R$ and mass $m$ will sweep up pebbles at a rate $dm/dt = \pi R^2 \rho_\mathrm{peb} \Delta v$, where $\rho_\mathrm{peb}$ is the mass density of pebbles and $\Delta v$ the approach velocity. Defining $q=m/m_\star$ and $\sigma=\Delta v/v_K$ as the dimensionless mass and relative velocity with $v_K=r\Omega_K$ as the Keplerian velocity corresponding to orbital radius of $r$, we can rewrite the accretion rate as
\begin{equation}
	\frac{dq}{dt} = A_\mathrm{geo} \frac{\chi_\bullet^2 \sigma q^{2/3}}{t_0},
\end{equation}
where $\chi_\bullet$ is the physical-to-Hill radius ratio
\begin{equation}
	\label{eq:chi}
	\chi_\bullet = \frac{R}{r} \left(\frac{3m_\star}{m}\right)^{1/3}
	= \left( \frac{9m_\star}{4\pi \rho_\bullet r^3} \right)^{1/3} \approx 10^{-4}-10^{-3},
\end{equation}
where $\rho_\bullet$ the internal density of the planetary body and $A_\mathrm{geo} = \pi/3^{2/3}\approx1.5$. In addition, the characteristic growth timescale is
\begin{align}
	\label{eq:T0}
	t_0 & = \frac{M_\odot}{\rho_\mathrm{peb} r^3 \Omega}                                                                                                                                                                  \\
	    & = 3.5\times10^4\,\mathrm{yr} \left( \frac{Z_\mathrm{peb}}{0.01} \right)^{-1} \left( \frac{\Sigma_\mathrm{gas}r^2/M_\odot}{0.01} \right)^{-1} \frac{H_\mathrm{gas}/r}{0.07} \frac{\Omega^{-1}}{20\,\mathrm{yr}},
\end{align}
where $\rho_\mathrm{peb}=Z_\mathrm{peb}\rho_\mathrm{gas} = Z_\mathrm{peb}\Sigma_\mathrm{gas}/\sqrt{2\pi}H_\mathrm{gas}$ the mass density of pebbles in the midplane, $Z_\mathrm{peb}$ the corresponding local metallicity, $\Sigma_\mathrm{gas}$ the surface density of the gas and $H_\mathrm{gas}$ the gas pressure scale height. The disk aspect ratio $h=H_\mathrm{gas}/r$ can also be written
\begin{equation}
	\label{eq:hgas}
	h = \sqrt{\frac{k_B T r}{\mu m_\mathrm{u} G m_\odot}}
	= 0.069 \left( \frac{T}{60\,\mathrm{K}} \right)^{1/2} \left( \frac{r}{\mathrm{20\,au}}\right)^{1/2},
\end{equation}
where $T$ is the temperature in the midplane and $\mu=2.34m_\mathrm{u}$ the mean molecular weight of the gas.

The degree of coupling between particles and gas is quantified by the particle stopping time $t_\mathrm{stop}$, which represents the time for a particle to adjust its motion to that of gas. Its dimensionless equivalent, $\tau_s=\Omega_K t_\mathrm{stop}$, is referred to as the aerodynamic size (in the literature, the Stokes number or St, is also common). In this work we define pebbles as particles of $\tau_s<1$. For the Epstein drag, $\tau_s$ can be expressed as $\tau_s = \pi (\rho_\bullet R)_\mathrm{peb} /2\Sigma_\mathrm{gas}$, indicating that the aerodynamic size is proportional to $R_\mathrm{peb}$.

The relative velocity $\Delta v = \sigma v_K$ can be identified in one of three ways:
\begin{itemize}
	\item the eccentricity of planetesimals, $\Delta v \simeq ev_K$ ($\sigma = e$), which is not considered in this work;
	\item the headwind velocity of the gas. Due to the radial pressure gradient, the azimuthal motion of the gas lags the Keplerian motion by an amount $\Delta v = \eta v_K$ ($\sigma=\eta$), where
	\begin{equation}
		\label{eq:eta}
		\eta \equiv -\frac{1}{2} \frac{\partial \log P}{\partial \log r} h^2
		= 4.9\times10^{-3} \left( \frac{\partial_{\log r} \log P}{-2} \right) \left( \frac{h}{0.07} \right)^2
	\end{equation}
	\citep[e.g.,][]{Weidenschilling1977}
	is a parameterization of the radial pressure gradient. Pebbles with $\tau_s<1$ follow the gas, while planetesimals ($\tau_s\gg 1$) move at the Keplerian velocity. Hence, the pebble-planetesimal velocity is also $\eta v_K$;
	\item the turbulent velocity $\Delta v \simeq \mathcal{M} c_s$ ($\sigma=\mathcal{M}h$), where $\mathcal{M}$ is the Mach number of the turbulent flow and $h$ is the disk aspect ratio.
\end{itemize}
Roughly, the highest of these velocities ($e$, $\eta$, or $\mathcal{M}h$) determines $\Delta v$.

Gravitational focusing increases the accretion rate by a factor of $(v_\mathrm{esc}/\Delta v)^2 = 2Gm/R(\Delta v)^2$, where $v_\mathrm{esc}$ is the escape velocity from its surface. In dimensionless units, the rate becomes
\begin{equation}
	\frac{dq}{dt} = A_\mathrm{Saf} \frac{\chi q^{4/3}}{(\sigma+q^{1/3}) T},
\end{equation}
with $A_\mathrm{Saf} = 2\pi/3^{1/3} \approx 4.4$. Here, we add a term $q^{1/3}$ in the denominator in order to account for the Keplerian shear component of the relative velocity, which becomes important when $\sigma$ is low.

Gravitational focusing significantly enhances the growth rates of planetesimals. On the other hand, its efficacy is hampered by the more copious noncollisional scatterings of planetesimals -- a process known as viscous stirring \citep{IdaMakino1993} -- or turbulent stirring \citep{IdaEtal2008}, which increases $\Delta v$. For small pebbles tightly coupled to the gas, this is not a concern, because gas drag circularizes their orbits before the next encounter. Therefore, in contrast to planetesimals, there is no need to follow the evolution of $\Delta v$ with time.

More potently, pebbles accrete by the settling mechanism, which is better known as pebble accretion \citep{OrmelKlahr2010,LambrechtsJohansen2012}. In pebble accretion, pebbles can be captured at distances much larger than $R$, after which the pebbles settle down in the gravitational potential. We adopt the numerically calibrated expression by \citet{OrmelLiu2018}\footnote{\citet{OrmelLiu2018} express these expression in terms of an efficiency $\epsilon = \dot{M}/2\pi r \Sigma v_r$. We take their 3D expression, $\epsilon = 0.39 q /\eta h_\mathrm{peb} \times f_\mathrm{set}^2$. With the drift velocity $v_r = 2\eta v_K \tau_s$ and $\rho_\mathrm{peb}=\Sigma_\mathrm{peb}/\sqrt{2\pi}h_\mathrm{peb}$, we obtain \eq{rate-settling}.}:
\begin{equation}
	\label{eq:rate-settling}
	\frac{dq}{dt} = A_\mathrm{peb} \frac{q \tau_s}{t_0} f_\mathrm{set}^2.
\end{equation}
Here $A_\mathrm{peb}=12.3$ and $f_\mathrm{set} \le 1$ is a modulation factor that suppresses accretion rates at low masses. It is a defining characteristic of pebble accretion that rates do not depend on the physical radius (or the nondimensional $\chi_\bullet$ parameter; \citealt{Ormel2017}). In addition, we assumed the 3D accretion regime for pebbles in \eq{rate-settling}, for which $\sigma$ does not enter explicitly. The velocity matters in the modulation factor $f_\mathrm{set}$, however:
\begin{equation}
	\label{eq:fset2}
	f_\mathrm{set}^2 = \exp \left[ - \left(\frac{\Delta v}{v_\ast} \right)^2 \right]
	\quad\textrm{or}\quad
	f_\mathrm{set}^2 = \left[ 1 + a_\mathrm{turb} \left( \frac{\Delta v}{v_\ast} \right)^2 \right]^{-3}.
\end{equation}
Here $v_\ast = (q_p/\tau_s)^{1/3} v_K$ and $a_\mathrm{turb}=0.33$ \citep{LiuOrmel2018}. The first expression in \eq{fset2} holds for a laminar-dominated velocity field and the second for a turbulent-dominated $\Delta v$. In the case of a turbulent flow, $\Delta v$ should be regarded as the rms value of an underlying velocity distribution that is isotropic \citep{OrmelLiu2018}.

\begin{table*}[tb]
	\renewcommand{\arraystretch}{1.1}
	\centering
	\caption{\label{tab:model-pars}Model parameters.}
	\small
	\begin{tabular}{llllp{6cm}}
		\hline
		Description                               & Symbol                & Unit         & Value/Range								& Reference/comment                                              \\
		\hline
		second component mass fraction				& $f_2$					&			& $\mathcal{U}(0,0.3)$					& only in model featuring a second particle size \\
		mass in accreted pebbles                  & $m_\mathrm{peb}$      & [$m_\oplus$] & $\mathcal{L}(1,10^3)$ & includes only the mass of pebbles that are consumed            \\
		primordial belt characteristic mass       & $m_\mathrm{brk}$      & [g]          & $\mathcal{L}(2\times10^{19},2\times10^{22})$                                                                  \\
		mass in primordial belt (planetesimals)   & $m_\mathrm{imf}$      & [$m_\oplus$] & $\mathcal{L}(0.3,105)$                                                                  \\
		pebble aerodynamic size (Stokes number) & $\tau_s$              &              & $\mathcal{L}(10^{-4},1)$      &                                                                \\ 
		second pebble aerodynamic size	  & $\tau_2$				&			& $\mathcal{L}(10^{-4},1)$					& only in model featuring a second particle size \\
		pebble relative velocity                  & $\sigma = \Delta v/v_K$ &            & $\mathcal{L}(4\times10^{-5},0.1)$		& relative velocity between bodies and pebbles                                           \\
		-- headwind parameter                     & $\sigma=\eta$         &              &											& see \eq{eta}                                       \\
		-- turbulent Mach number                  & $\sigma=\mathcal{M}h$ &              &											& $\Delta v$ follows a distribution; $h$ is the aspect ratio of the disk and $\mathcal{M}$ the turbulence Mach number.                          \\
		\hline
		internal density parameter                & $\chi$                &              & $5\times10^{-4}$         & \eq{chi}. Irrelevant for pebble accretion.                     \\
		implantation factor                       & $f_\mathrm{impl}$     &              & $2\times10^{-3}$         & fraction of bodies in the primordial belt that ends up in P48. \\
		\hline
	\end{tabular}
\end{table*}

\begin{figure}
	\includegraphics[width=\columnwidth]{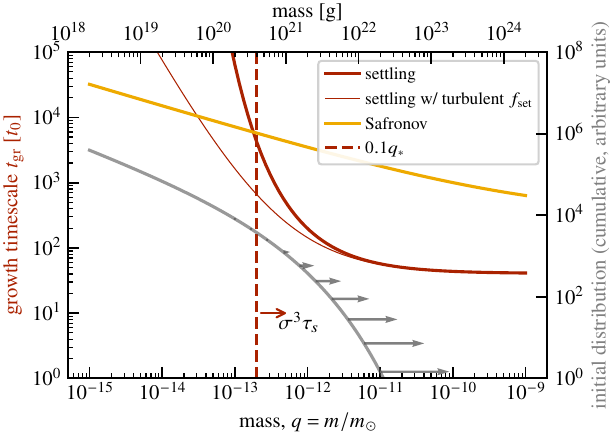}
	\caption{\label{fig:timescales}Growth timescales for pebble accretion (settling mechanism, red) and Safronov accretion (gravitational focusing in the absence of gas drag; yellow) for a velocity parameter of $\sigma=\Delta v/v_K = 10^{-3}$, dimensionless stopping time $\tau_s=2\times10^{-3}$, and internal density parameter $\chi_\bullet=5\times10^{-4}$. The critical mass $q_\ast = \sigma^3 \tau_s$ (or rather a fraction of this number; vertical dashed line) indicates the point where the settling mechanism becomes operational. In the case where the approach velocity is due to turbulence, such that $\Delta v$ follows a distribution, the transition is more gradual (thin red line). Pebble accretion acts on the high-mass tail of the distribution where the length of the arrows is proportional to the growth rate (\eq{rate-settling}; gray).
    }
\end{figure}

\Fg{timescales} presents the Safronov and settling growth rates as functions of mass. Here, the rates are expressed in terms of the growth timescale, $t_\mathrm{gr} = m/(dm/dt)$, which is given in units of $t_0$ (see \eqr{T0}).  At low masses, accretion is dominated by gravitational focusing (Safronov limit; yellow curve). It is well known that in this limit, the growth timescale decreases with the mass: the runaway growth regime \citep{WetherillStewart1989}. Pebble accretion (the settling mechanism) is unimportant at low masses. However, at a mass scale that is a fraction of $q_\ast = \sigma^3 \tau_s$ (the mass corresponding to $\sigma \sim v_\ast$), the settling mechanism starts to become efficient, increasing the growth rate significantly \citep{VisserOrmel2016}. In the case of a turbulence-dominant velocity field, this transition is slightly less abrupt (thin line). Hence, bodies more massive than ${\sim}0.1\,m_\ast$ will undergo rapid growth, while those with $m\ll m_\ast$ grow more slowly. Pebble accretion therefore has the potential to flatten the high-mass tail of the size distribution, illustrated in \fg{timescales} by the gray arrows.

\subsection{Model parameters and likelihood}
\label{sec:parameters}
An overview of the parameters used in the model is given in \Tb{model-pars}. The initial distribution of bodies is specified by the parameter $m_\mathrm{brk}$ and the initial mass of the belt $m_\mathrm{imf}$  (see \eqr{Kavelaars2021}), which are uniformly sampled in the log space. The pebble properties are given by their dimensionless velocity $\sigma$, the dimensionless stopping time $\tau_s$, and the total accreted pebbles mass $m_\mathrm{Peb}$. In addition to the continuous sampling of these parameters, the model also employs a binary switch for the nature of the velocity field, either laminar or turbulent. If the velocity is laminar, $\sigma$ is identified with the normalized pressure gradient parameter $\eta$; if the velocity is turbulent, $\sigma$ is identified with the Mach number of the turbulent velocity field. The velocity field matters in calculating the $f_\mathrm{set}$ modulation factors (see \eqr{fset2}).

In total, the sampling provides us with several hundred bodies, which covers the entire mass distribution. The mass of each of the bodies is integrated over time. We work in dimensionless units where time is normalized by $t_0$ (\eqr{T0}) and mass by the mass of the Sun. The pebble accretion rate $dq/dt$ is computed for each of these bodies, according to the collision rates given in \se{rates}. The bodies do not interact among themselves, and because $dq/dt$ is an increasing function of $q$, the initial order of their masses stays preserved. Under these conditions, the calculation boils down to a set of ordinary differential equations, which we solve with the \texttt{scipy.integrate.odeint} library. Simultaneously with the increase in $q$, we decrease the mass of the pebble  reservoir $q_\mathrm{peb}$, whose initial value is also a parameter. When $q_\mathrm{peb}$ reaches 0, we stop the simulation and record the elapsed time $t$ (in units of $t_0$). This distribution is then weighted by the implantation factor $f_\mathrm{impl}$, the ratio of the number of bodies in the present-day P48 to that in the primordial belt, to obtain the simulated distribution of bodies in P48.

\begin{table*}
	\centering
	\small
	\renewcommand{\arraystretch}{1.3}
	\caption{\label{tab:outputtable}List of model runs and posterior values for the MCMC simulations.} 
\begin{tabular}{llllllllll}
\hline
\hline
model  & $\log_{10} \sigma$ & $\log_{10} \tau_s$ & $\log_{10} \tau_2$ & $f_2$ & $m_\mathrm{brk}$ [$10^{20}\,\mathrm{g}$] & $m_\mathrm{imf}$ [$m_\oplus$] & $m_\mathrm{peb}$ [$m_\oplus$] & $C_{\tau\sigma}$ & $\ln \mathcal{P}$ \\
\hline
\texttt{lami-1p}                    & $-2.78^{+0.40}_{-0.38}$ & $-1.70^{+1.15}_{-1.22}$ &  &  & $8.90^{+2.12}_{-1.96}$ & $11.61^{+1.00}_{-1.08}$ & $5.22^{+4.09}_{-2.25}$ & $-10.05^{+0.11}_{-0.12}$ & $-86.64$ \\
\texttt{turb-1p}                    & $-2.56^{+0.41}_{-0.49}$ & $-2.12^{+1.46}_{-1.27}$ &  &  & $6.69^{+1.96}_{-1.49}$ & $11.39^{+1.24}_{-2.12}$ & $7.60^{+5.45}_{-3.61}$ & $-9.81^{+0.13}_{-0.13}$ & $-86.14$ \\
\texttt{lami-2p}                    & $-2.92^{+0.43}_{-0.30}$ & $-1.88^{+1.27}_{-1.36}$ & $-1.96^{+1.32}_{-1.33}$ & $0.14^{+0.11}_{-0.10}$ & $8.14^{+2.20}_{-2.36}$ & $11.61^{+1.02}_{-1.09}$ & $5.60^{+4.48}_{-2.47}$ & $-10.12^{+0.15}_{-1.74}$ & $-86.10$ \\
\texttt{turb-2p}                    & $-2.84^{+0.52}_{-0.30}$ & $-1.88^{+1.32}_{-1.48}$ & $-2.30^{+1.58}_{-1.19}$ & $0.15^{+0.10}_{-0.10}$ & $6.21^{+1.86}_{-1.57}$ & $11.66^{+1.14}_{-1.34}$ & $7.10^{+5.30}_{-3.28}$ & $-9.86^{+0.16}_{-0.62}$ & $-85.96$ \\
\hline
\end{tabular}
	\tablefoot{See \Tb{model-pars} for the parameters and \eq{tausig} for $C_{\tau\sigma}$. The lower and upper values indicate the 16\% and 84\% posterior ranges. The last column gives the maximum likelihood of the sample.}
\end{table*}

The likelihood function is calculated by comparing the simulated distribution with the reconstructed distribution in the \ossp domain and the observed number of bodies in the MPC domain (see \fg{cumulative}). To this end, the particles are binned by mass, with 10 bins covering the \ossp region and 15 covering the MPC region; the bin width is about 0.2 dex. The number of bodies in the synthetic and observed distributions are then compared for each bin. In the \ossp region, Gaussian statistics are assumed; in other words, the contribution to the log-likelihood of bin $k$ equals
\begin{equation}
	\label{eq:logli-ossos}
	\ln P_k^\mathrm{\ossp} = -\frac{1}{2} \left( \frac{N_k -N_{\mathrm{sim},k}}{\sigma_k} \right)^2 -\frac{1}{2} \ln (2\pi \sigma_k^2)
\end{equation}
where $N_{\mathrm{sim},k}$ is the simulated number of bodies, $N_k$ is the observed number of bodies in P48, as obtained from \eq{dNdHfit}, and $\sigma_k$ represents the spread. The spread $\sigma_k$ is also obtained from the debiased \ossp distribution.\footnote{We estimate the 2$\sigma$ error, again from inspecting Figure~1 of \citep{PetitEtal2023}, to be decreasing linearly with $H_r$ from 0.4 dex at $H_r=5$ to 0.13 dex at $H_r=6.8$, whereafter it stays constant at this value.}
For the MPC part, we use binomial statistics to obtain the likelihood contribution:
\begin{equation}
	\label{eq:logli-mpc}
	\ln P_k^\mathrm{MPC} = \ln \mathcal{B}(N_k,N_{\mathrm{sim},k},f_\mathrm{impl}).
\end{equation}
Here $\mathcal{B}(x,n,p)$ is the probability of drawing $x$ successes out of $n$ independent samples where each draw has probability $p$.

In addition, we apply two priors to the likelihood function, related to the following observations. 
First, the number of Jupiter trojans constrains the total mass of the primordial belt $m_\mathrm{imf}+m_\mathrm{peb}$. Since the Jovian trojans are also thought to be captured from the same disk, the mass of bodies present in the primordial belt can be estimated using a numerically simulated  implantation rate and the observed mass of the trojans (see Figure~\ref{fig:P48} and \citealt{Nesvorny.2013}). The total pre-instability mass of the primordial belt falls in the range of $14{-}28\,m_\oplus$. Consequently, we add a term $\mathcal{P}^\mathrm{PB}$ to the likelihood function corresponding to a situation where the parameter $\mu = \ln(m_\mathrm{imf}+m_\mathrm{Peb})$ is normally distributed, centered around $\ln 20\,m_\oplus$ and with spread $\sigma=0.34$. Second, the preservation of the dynamically cold classicals limits the mass of individual bodies. If any planetary-mass bodies formed in the primordial outer disk and were scattered out by a migrating Neptune, the cold classical belt would likely be overheated or even destroyed by the gravitational sweeping of such a planet. Therefore, the masses of the individual bodies making up the primordial belt could not have been too large. \citet{PetitEtal1999} simulated the effect of massive protoplanets on the stability of the cold belt, and found a threshold mass of ${\sim}1\, m_\oplus$. To incorporate this mass constraint, we add a term $\ln \mathcal{P}^\mathrm{CC} = -(m_1/m_\oplus)^2$ to the log-likelihood function, which suppresses models favoring the formation of bodies beyond 1 Earth mass. 
The total log-likelihood is then
\begin{equation}
	\ln \mathcal{P}
	= \sum_k \ln \mathcal{P}_k^\mathrm{MPC}
	+ \sum_k \ln \mathcal{P}_k^\mathrm{MPC}
	+ \ln \mathcal{P}^\mathrm{CC}
	+ \ln \mathcal{P}^\mathrm{PB}.
\end{equation}
We conduct an MCMC analysis using \textit{emcee} \citep{Foreman-MackeyEtal2013}, marginalizing over the input parameters listed in \Tb{model-pars}.

\begin{figure*}
	\centering
	\includegraphics[width=0.9\textwidth]{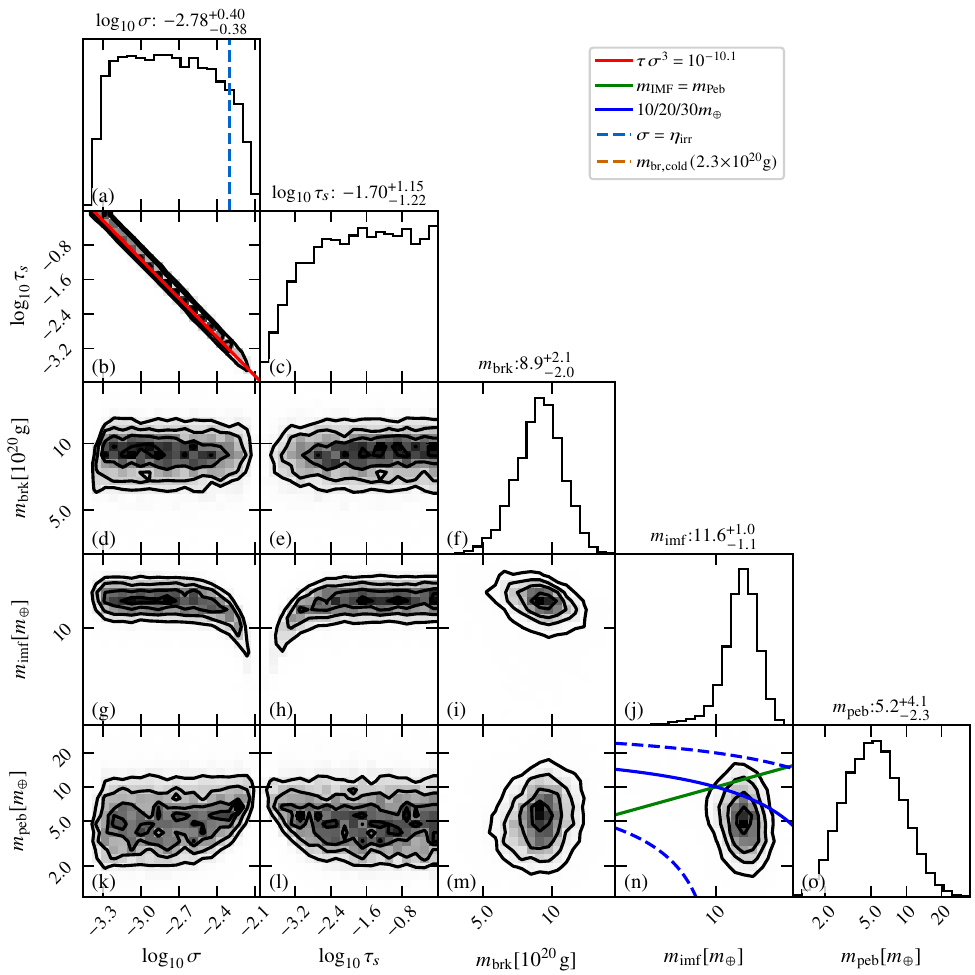}
	\caption{\label{fig:corner-lam-1p}MCMC posterior distribution of the single particle laminar model (\texttt{lami-1p}). The 16\%, 50\%, and 84\% confidence levels are indicated above each histogram. The runs cluster around the line of constant $\tau_s \sigma^3$, which indicates a mass scale above which pebble accretion operates (panel b). High-velocity solutions, as in a smooth disk ($\sigma=\eta_\mathrm{irr}$), have difficulty fitting the data (panel a). The mass of the accreted pebbles is lower than the mass of the initial popluation by around 0.4 dex (panel n).}
\end{figure*}
\section{Results}
\label{sec:results}
The posterior values of the model parameters from the MCMC fit are summarized in \Tb{outputtable}. Four model suites are considered, each characterized by binary choices for the velocity field---either laminar (\texttt{lami-}) or turbulent (\texttt{turb-})---and the particle components---a single particle size (\texttt{1p}) or two pebble sizes (\texttt{2p}). Models with two particle sizes feature two additional parameters: the aerodynamic size ($\tau_2$) and the mass fraction of the second component ($f_2$).

\subsection{Simulations with a single particle component}
\Fg{corner-lam-1p} presents the corner plot of the \texttt{lami-1p} model (laminar velocity field with one pebble size). The most striking feature is the correlation between the stopping time $\tau_s$ and the relative velocity (see panel b). The slope of this correlation is $-3$, which points to a mass scale where pebble accretion is initiated, $q_\ast = \sigma^3 \tau_s$. Specifically, we fit
\begin{equation}
	\label{eq:tausig}
	C_{\tau\sigma} \equiv \log_{10} \tau_s \sigma^3 =
	\log_{10} \tau_s +3\log_{10} \left( \frac{\Delta v}{v_K} \right)
	= -10.05^{+0.11}_{-0.12}.
\end{equation}
In physical units, $m_\ast = 10^{C_{\tau\sigma}} m_\odot \approx 1.6\times10^{23}\,\mathrm{g}$  (or $10^{-5}\,m_\oplus$; the actual mass where pebble accretion starts to dominate over Safronov accretion is ${\approx}10\%$ of this number; see \fg{timescales}). For bodies of mass $m>m_\ast$, pebble accretion is fully operational, whereas below $m_\ast$ pebble accretion is suppressed. Therefore, for $m \ll m_\ast$, the mass distribution tends to follow the IMF, as given by the $m_\mathrm{brk}$ and $m_\mathrm{imf}$ parameters.
However, due to the exponential tapering at high mass, the IMF cannot fit the high-mass end of the P48 bodies. Here pebble accretion acts to flatten the mass distribution (see \fg{timescales}). \Tb{outputtable} shows that about $5{-}8$ Earth masses in pebbles are needed, an amount that is similar to the mass in the IMF.

\begin{table*}
	\centering
	\small
	\renewcommand{\arraystretch}{1.25}
	\caption{\label{tab:runpars}Parameters of several runs presented in \fg{sim-default}.}
\begin{tabular}{lllllllllll}
\hline
\hline
name & $\sigma$ & $\tau_s$ & $\tau_2$ & $f_2$ & $m_\mathrm{brk}$ & $m_\mathrm{imf}$ & $m_\mathrm{peb}$ & mode & time & $\Delta \ln \mathcal{P}$\\
 &  &  &  &  & [$10^{20}\,\mathrm{g}$] & [$m_\oplus$] & [$m_\oplus$] &  & [$t_0$] & \\ \hline
\hline
\texttt{lami-1p-best}                    & $1.1{\times}10^{-3}$ & $0.074$ &  &  & $8.7$ & $12$ & $7.8$ & L & $9.1$ & $0.0$\\
\texttt{lami-1p-etairr}                  & $6.4{\times}10^{-3}$ & $3.4{\times}10^{-4}$ &  &  & $10$ & $10$ & $8.4$ & L & $1.8{\times}10^{3}$ & $-1.2$\\
\texttt{lami-1p-himass}                  & $1.1{\times}10^{-3}$ & $0.092$ &  &  & $11$ & $11$ & $11$ & L & $7.6$ & $-1.1$\\
\texttt{turb-1p-best}                    & $9.7{\times}10^{-3}$ & $1.4{\times}10^{-4}$ &  &  & $10$ & $6.1$ & $12$ & T & $4.6{\times}10^{3}$ & $0.50$\\
\texttt{turb-1p-hitau}                   & $1.1{\times}10^{-3}$ & $0.16$ &  &  & $8.1$ & $11$ & $13$ & T & $5.5$ & $-0.68$\\
\texttt{lami-2p-best}                    & $5.1{\times}10^{-3}$ & $4.0{\times}10^{-4}$ & $5.2{\times}10^{-3}$ & $0.085$ & $7.4$ & $10$ & $8.3$ & L & $1.2{\times}10^{3}$ & $0.54$\\
\texttt{turb-2p-best}                    & $8.6{\times}10^{-3}$ & $1.8{\times}10^{-4}$ & $1.9{\times}10^{-4}$ & $0.25$ & $7.2$ & $7.6$ & $11$ & T & $3.8{\times}10^{3}$ & $0.68$\\
\hline
\end{tabular}
	\tablefoot{The last two columns denote the time (in units of $t_0$) to complete the run and the change in log-likelihood with respect to the \texttt{lami-1p-best} model.}
\end{table*}
\begin{figure*}
	\centering
	\includegraphics[width=\textwidth]{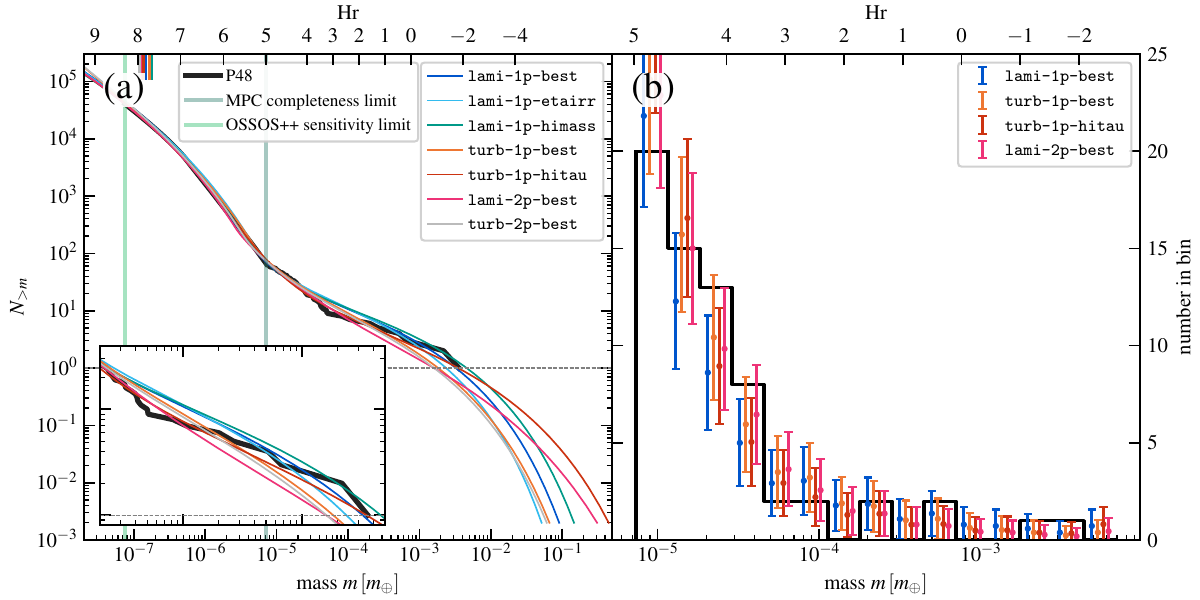}
	\caption{\label{fig:sim-default}Cumulative distributions (a) and histograms (b) of sample model runs. An implantation factor of $f_\mathrm{impl}=2\times10^{-3}$ has been applied, after which the simulated distributions can be compared to the P48 population (black). In (b) the low number of bodies (and large Poisson error bars) illustrate why the fits tend to be rather insensitive to the specifics at the high-mass end of the distribution. Runs featuring small $\tau_s$ are associated with high $\sigma$ due to the degeneracy between these quantities (see \Tb{runpars}). Runs with small $\tau_s$ are disfavored, however, as pebble accretion operates too slowly.
	}
\end{figure*}
\begin{figure}
	\includegraphics[width=\columnwidth]{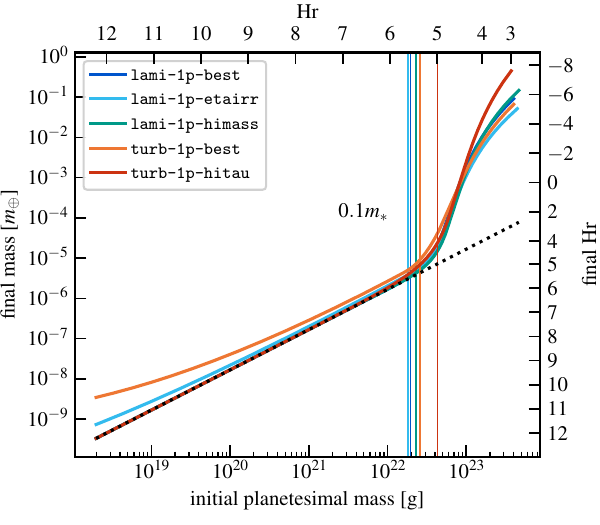}
	\caption{\label{fig:massgain}Final \textit{vs} initial mass for several simulation runs (see \Tb{runpars}). The vertical lines denote 10\% of $m_\ast = \sigma^3 \tau_s m_\odot$, which is the scale where pebble accretion begins. The dashed line is the identity line ($y=x$). Bodies ${\gtrsim}10^{22}$ grow their mass by several tenfold by pebble accretion and become dwarf planets. The gain in mass at low mass in the \texttt{lami-1p-etairr} and \texttt{turb-1p-best} is due to the slow pace of pebble accretion, allowing Safronov interactions to become important at low planetesimal masses.
	}
\end{figure}

At high velocity the correlation starts to break down.
Here, the combination of high $\sigma$ and small $\tau_s$ renders pebble accretion very slow (\eq{rate-settling}) and allows ballistic encounters (Safronov focusing) to rival pebble accretion. As a result, the mass distribution at low mass is altered, which is demonstrated by the \texttt{lami-1p-etairr} model, which features a value of $\sigma$ consistent with that of an irradiated disk.
\Fg{massgain} illustrates that in this model low-mass bodies accrete a significant amount of pebbles. As $\sigma$ increases, this trend becomes more pronounced and results in poorer fits because the \ossp part of the distribution can no longer be matched. This finding is unsurprising as the shape of the IMF was adopted from the cold classicals, which was already shown to fit the shape of hot classicals across the \ossp range \citep{KavelaarsEtal2021,PetitEtal2023}. However, it also shows that accretion processes cannot maintain the shape of the low-mass part, at least not in conjunction also matching the high-mass tail. It lends further support to the primordial nature of the low-mass part of the TNO size distribution.

An additional concern for these small $\tau_s$ runs is that they take a long time ($t/t_0$ is high). We can estimate the characteristic accretion time for pebble accretion from the pebble accretion rate (\eq{rate-settling}): $t\sim q/\dot{q} \sim t_0/\tau_s$, for $f_\mathrm{set}\sim1$ (i.e., $m>m_\ast$). This estimate agrees with \Tb{runpars} where we compiled $t/t_0$ for selected model runs. We recall that the simulations are conducted in dimensionless parameter space, where we can hide our ignorance of, the pebble density, for example, which sets the value of $t_0$ (\eqr{T0}). However, the pebble accretion mechanism requires gas-rich conditions and it must terminate \textit{before} the lifetime of the gas-rich disk, typically estimated at ${\sim}5\,\mathrm{Myr}$ \citep{WeissEtal2021}.
Therefore, a high $t/t_0$ renders these models implausible. We return to the timescale constraints in \se{scenarios}.

\Tb{outputtable} shows that a turbulent velocity field marginally improves the log-likelihood of the best model by an amount of +0.5. This specific run is characterized by a very low $\tau_s$ value (\texttt{turb-1p-best} entry in \Tb{runpars}), which is disfavored in light of the above discussion. A higher $\tau_s$ run, following the $\tau_s$--$\sigma$ correlation (\texttt{turb-1p-hitau}) fits the data equally well. \Fg{sim-default}a presents the cumulative distribution of several individual runs. All models overshoot the infliction point 
at $m=10^{-4}\,m_\oplus$ and also tend to undershoot the final bin where the observed CMF terminates (with Triton) at $m=3\times10^{-3}\,m_\oplus$.
These apparent mismatches are a consequence of the low number of TNOs at these masses, which can be understood from inspecting 
the differential distribution (on which the likelihood calculations are based). \Fg{sim-default}b presents these fits for the MPC part of the distribution. Here, the error bars denote Poisson counting uncertainties. The bin to the right of $m=10^{-4}\,m_\oplus$ with 0 bodies in the MPC is typically fitted with ${\approx}2$ bodies in the simulation, but this still has a Poisson likelihood of 13.5\%. Similarly, the simulation runs are quite insensitive to the presence of the two most massive bodies (Pluto and Triton); statistically, the low number of high-$m$ bodies do not carry much weight. As a result the runs that undershoot the cumulative curve in \fg{sim-default}a (e.g., \texttt{turb-1p-best}) statistically cannot be distinguished from those runs that include a larger $\tau_s$ (e.g., \texttt{turb-1p-hitau}) and those runs that contain more pebbles (e.g., \texttt{turp-1p-himass}). 

\subsection{A second particle component}

The addition of a secondary component does not significantly improve the results.  From \Tb{outputtable}, the Bayes factor (the difference in $\ln \mathcal{P}$) is $0.54$ between the laminar models and $0.18$ between the turbulence models. These values indicate that the added parameters have insignificant effects on the fits \citep{Jeffreys1998}. This insensitivity is reflected in \Tb{outputtable} by the proximity of the size of the secondary component ($\tau_2$) to that of the main component ($\tau_s$) in the \texttt{lami-2p} and \texttt{turb-2p} models. \Fg{sim-default} presents this result graphically. The \texttt{lami-2p-best} run---the run where $\mathcal{P}$ is the highest---hardly stands out from the other runs. Curiously, these models tend to be fitted by low $\tau_s$ (and $\tau_2$) and correspondingly high $t_0$, indicating that these runs improve the fit for the \ossp part of the distribution, rather than the high-mass MPC part. In other words, the low number of high-mass bodies in P48 prevents any meaningful fit for the high-mass end of the distribution, which makes the model insensitive to a second particle component.

While we do not find evidence for a binary distribution of pebbles, two points should be noted. First, the size distribution could still be a power law, which would ameliorate the $f_\mathrm{set}$ transition in a similar way to that of the turbulence velocity field \citep{LyraEtal2023}. Second, the lack of evidence for a second component does not preclude its presence. A bigger sample of high-mass bodies is necessary to establish or discount its existence.

\section{Implications}
\label{sec:scenarios}
The MCMC calculations in the previous section demonstrate that both turbulent and laminar models fit the data. Here, we identify the laminar model with formation in a standard smooth disk, characterized by a spatially constant radial pressure parameter $\eta$ (\eq{eta}). Conversely, the turbulent model corresponds to formation within an environment where pebbles collect around a radial pressure maximum ($\eta\approx0$) such that velocities are dominated by turbulent motions. This can be identified with pebble rings seen in ALMA.

\subsection{Formation in a pebble ring}
Pebble rings are ubiquitous in ALMA imagery \citep{AndrewsEtal2018,BaeEtal2023}. The containment of particles by pressure maxima is the leading explanation for the observed particles accumulation of particles in these rings. Around the pressure maximum location $\eta\approx0$ and the relative motion between pebbles and bigger bodies is due to turbulence. Studies have used the extent of rings in both the vertical and radial dimensions to constrain the diffusivity (the ``turbulent alpha'' of the gas). Specifically, this constrains the ratio of gas diffusivity over Stokes number of the particles, $\alpha_D/\tau_s$. \citet{Rosotti2023} compiled results from a number of works that measure the radial width of continuum rings (their Table 2), and found that the inferred $\alpha_D/\tau_s$ ranged from a low of 0.04 to values exceeding unity (the latter essentially implies that particles follow the pressure distribution). However, $\alpha_D/\tau_s$ obtained from the vertical extent of the dust layer are often inferred to be smaller, e.g., $\alpha_D/\tau_s\approx10^{-2}$ \citep{VillenaveEtal2022}. For simplicity, we adopt in the following a range in $\alpha_D/\tau_s$ between $0.01$ and $0.4$, which covers most of these measurements.

When we convert the diffusivity parameter $\alpha_D$ to a root mean square (rms) particle velocity, we obtain\footnote{Here we have related the parameter $\alpha_D$, which expresses the diffusivity of the gas, to the rms velocity of the turbulent gas motions $\sigma$. The turbulent diffusivity of the gas is parameterized in terms of $\alpha$ as $D_\mathrm{gas} = \alpha_D c_s^2 /\Omega$. In general, we can write $D_\mathrm{gas} = V_L^2 t_L$, where $V_L$ and $t_L$ are, respectively, the velocities and turnover times of the largest eddies in the turbulent cascade. A common choice for rotating systems is $t_L=\Omega_K^{-1}$ and therefore $V_L = \sqrt{\alpha_D} c_s$ \citep{CuzziEtal2001}. Furthermore, if the turbulence is Kolmogorov, the rms velocities of the turbulent gas is $\Delta v = \mathcal{M}c_s = (3/2)V_L$. Then, we obtain $\alpha_D = (2\mathcal{M}/3)^2 = (2\sigma/3h)^2$.}
\begin{equation}
	\frac{\sigma^2}{\tau_s} = \frac{3h^2}{2} \left( \frac{\alpha_D}{\tau_s} \right)_\mathrm{ALMA}
	\approx 7\times10^{-4} \left( \frac{h}{0.07} \right)^2 \left( \frac{\alpha_D/\tau_s}{0.1} \right)_\mathrm{ALMA}
	\label{eq:constr1}
\end{equation}
where $(\alpha_D/\tau_s)_\mathrm{ALMA}$ is inferred from the extent of the ring in ALMA imagery, and $h=0.07$ is our adopted value for the aspect ratio of the gas at the location of the primordial belt.

\begin{figure}
	\includegraphics[width=\columnwidth]{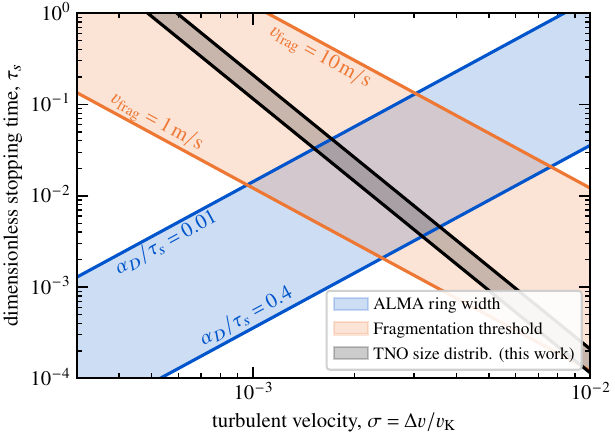}
	\caption{\label{fig:tausig}Constraints on the particle aerodynamic size (Stokes number $\tau_s$) and turbulence intensity of the gas, where $\sigma$ is identified as the root mean square turbulent gas velocity.  The constraints follow from simulating the TNO size distribution (run \texttt{turb-1p}, black line), ALMA ring morphological analysis (blue lines; \citealt{Rosotti2023}), and particle fragmentation velocity (orange lines; the adopted value for the Keplerian velocity $v_K$ is the circular motion at 20 au). Under the assumption that the primordial belt featured conditions similar to ALMA rings, the degeneracy between $\tau_s$ and $\sigma$ (or $\alpha_D$; \eq{constr1}) can be lifted and we find, $\tau_s\sim0.01$, $\alpha_D \sim 10^{-3}$, and a fragmentation threshold of ${\sim}2\,\mathrm{m\,s}^{-1}$. }
\end{figure}
A further constraint to particle sizes in rings comes from the fragmentation barrier \citep{GuettlerEtal2010,BirnstielEtal2011}. When the gas is turbulent, the pebble-pebble relative velocity increases with the size (Stokes number) as $\Delta v_\mathrm{pp} \approx \mathcal{M} c_s \sqrt{2\tau_s}$ \citep{OrmelCuzzi2007},\footnote{This is the rms particle-particle relative velocity between two particles in the inertial range. The rms particle-planetesimal relative velocity for $\tau_s<1$ particles is $\Delta v = \mathcal{M} c_s$.} where $\mathcal{M}$ is the Mach number of the turbulent flow. The aerodynamic size is then determined by the condition 
$\Delta v_\mathrm{pp} = v_\mathrm{frag}$, resulting in
\begin{equation}
	\sigma \tau_s^{1/2}
	= \frac{\mathcal{M} c_s}{v_K} \tau_s^{1/2}
	= \frac{\Delta v_\mathrm{pp}}{2^{1/2}v_K} 
	\lesssim 1.1\times10^{-4} \left( \frac{v_\mathrm{frag}}{1\,\mathrm{m\,s^{-1}}} \right).
\end{equation}
These constraints are presented in \fg{tausig}, together with the constraint on $\tau_s \sigma^3$ that we obtained from modeling the TNO size distribution (\Tb{outputtable}). The constraint on the extent of ALMA rings is almost orthogonal to the constraint from the TNO size distribution modeling (this study), allowing us to put meaningful constraints on $\tau_s$ and $\alpha_D$. Together, they point to an aerodynamic size in the range of $\tau_s \sim [3\times10^{-3}{-}4\times10^{-2}]$, a turbulent diffusivity of $\alpha_D\sim [4\times10^{-4}{-}2\times10^{-3}]$, and a fragmentation threshold of $v_\mathrm{frag}\sim [1.7{-}3.2\,\mathrm{m\,s}^{-1}]$.

These values are similar to what was found by \citet{JiangEtal2024} in a recent analysis, who assumed a fragmentation threshold of $v_\mathrm{frag}\sim1\,\mathrm{m\,s}^{-1}$. Our results independently suggest that the particle fragmentation threshold amount is ${\sim}2\,\mathrm{m\,s}^{-1}$ in order for particles to grow to the inferred aerodynamic sizes. A higher fragmentation threshold is possible if the growth of particles is limited by, for example, bouncing \citep{ZsomEtal2010}. A fragmentation velocity increase by a factor of 2 would increase the derived value of the fragmentation $\alpha$ in \citet{JiangEtal2024} by a factor of 4 to ${\sim}4\times10^{-4}$, in line with our findings.

If the TNO population formed from a ring in which the pebbles are contained by pressure, it is natural to assume that the birth ring would have had a mass in pebbles similar to the accreted pebble mass $m_\mathrm{peb} \sim 10m_\oplus$. This allows us to constrain the characteristic timescale $t_0$, \eq{T0}. Since $\rho_\mathrm{peb} \sim m_\mathrm{ring}/2\pi r W_r H_r$, where $W_r$ and $H_r$ are the width and height of the ring, we obtain
\begin{equation}
	t_0 \sim 4\times10^{3}\,\mathrm{yr} \ \left( \frac{m_\mathrm{ring}}{10\,m_\oplus} \right)^{-1} \frac{W_r/r}{0.1} \frac{H_r/r}{0.01} \frac{\Omega^{-1}}{20\,\mathrm{yr}}.
\end{equation}
If the pebble aerodynamic size is $\tau_s \sim 10^{-2}$, it would take a time ${\sim}t_0/\tau_s$ for these bodies to accrete the pebbles, i.e., ${\sim}0.4\,\mathrm{Myr}$, for the geometrical factors $W_r/r$ and $H_r/r$ adopted above. This time falls within the typical lifetimes of the solar nebula disk \citep{WeissEtal2021}, implying that pebbles will be consumed before the disk dissipates. However, very wide rings ($W_r/r \sim 1$ or $H_r/r \sim 0.1$) would dilute the concentration of pebbles and render pebble accretion much slower.

\subsection{A laminar disk}
If pebbles are not entrained by a pressure maximum and freely drift inward, the relative velocity between planetesimals and pebbles is given by the global disk pressure gradient $\eta$. From \eq{eta},
$\sigma = \eta \approx 5\times10^{-3}$ for a power-law index of pressure on radius of $2$ and from the $\tau_s$--$\sigma$ relationship (\eq{tausig}), we then find an aerodynamic size of a mere $\tau_s = 7\times10^{-4}$, which is rather small for pebbles. The radial drift timescale
\begin{align}
	\label{eq:tdrift}
	t_\mathrm{drift}
	 & = \frac{r}{v_r} = \frac{1}{2\eta\tau_s\Omega}                                                                                                                             \\
	 & = 2\times10^6\,\mathrm{yr}\left( \frac{\eta}{5\times10^{-3}} \right)^{-1} \left( \frac{\tau_s}{7\times10^{-4}} \right)^{-1} \left( \frac{r}{20\,\mathrm{au}}\right)^{3/2}
\end{align}
already rivals the disk lifetime, suggesting that particles hardly drift. In addition, the growth of planetesimals by such small particles would be excruciatingly slow, $t_\mathrm{gr}\sim t_0/\tau_s \sim 50\,\mathrm{Myr}$ for the parameters listed in \eq{T0}, far exceeding the lifetime of the gas disk. A growth timescale ${\lesssim}10^6$ yr would therefore imply a (midplane) dust-to-gas ratio $Z_\mathrm{peb}\sim1$, which seems unlikely for small $\tau_s$ particles. In summary, it is unlikely that TNOs could accrete pebbles from an inward-drifting pebble stream in the standard, laminar disk model.

In addition, pebble accretion in a laminar disk is lossy \citep{Ormel2017,LinEtal2018}. In the scenario where the birth ring is supported by pressure, all pebbles can (eventually) be accreted. That is, the efficiency is near 100\%. This contrasts with the smooth disk scenario, where pebbles that fail to be accreted are lost. Consequently, the total mass in pebbles \textit{required} for accretion, $m_\mathrm{needed}$, exceeds the mass in pebbles actually accreted ($m_\mathrm{peb}$). In a laminar disk, we can estimated $m_\mathrm{needed}$ as
\begin{align}
	\nonumber
	m_\mathrm{needed} & \sim 2\pi r v_r \rho_\mathrm{peb} H_\mathrm{peb}  t
	\sim 4\pi M_\odot \eta h \tau_s \sqrt{\frac{\alpha_z}{\alpha_z+\tau_s}} \frac{t}{t_0}                                                                \\
	                  & \sim 1.4\times10^3\,m_\oplus \left( \frac{h}{0.07} \right)^3 \left( \frac{\alpha_z/\tau_s}{1} \right)^{1/2} \frac{t}{t_0/\tau_s}
	\label{eq:m-needed}
\end{align}
where $t$ is the duration of the pebble accretion, $H_\mathrm{peb}=hr \sqrt{\alpha_z/(\alpha_z+\tau_s)}$ the pebble scale height \citep{DubrulleEtal1995}, and $\rho_\mathrm{peb}$ the pebble density for which we substituted \eq{T0}. In the second line, we used again $t\sim t_0/\tau$. \Eq{m-needed} highlights that a great number of pebbles---far exceeding the mass actually accreted ($m_\mathrm{peb}$)---is required, underscoring the inefficiency of pebble accretion in the primordial belt in a laminar setting. For small $\tau_s$, the problem is that pebbles are not sufficiently concentrated in the midplane, whereas large $\tau_s$ implies a short duration of pebbles in the primordial belt.

All these concerns would be reduced considerably when the disk is colder. Expressed in terms of the disk aspect ratio ($h\propto T^{1/2}$), $\eta \propto h^2$ and therefore $\tau_s \propto \sigma^{-3} \propto h^{-6}$ from \eq{tausig}. For example, when $h=0.05$ (about 35\,K at 20 au; see \eq{hgas}), $\tau_s \approx 6\times10^{-3}$, a factor of 10 increase, the drift timescale reduces by a factor of 4 compared to \eq{tdrift}, the growth timescale may fall below the lifetime of the disk, and the required pebble mass, $m_\mathrm{needed}$, to values below $500\,m_\oplus$. These numbers are within the realm of pebble accretion models \citep{LambrechtsEtal2019}, although on the high side. A question that arises, however, is where these $500\,m_\oplus$ in solids would end up as the Solar System's giant planets, while enriched, are not dominated in metals.

\section{Discussion}
\label{sec:discussion}
\subsection{Pebble accretion in the cold classical belt?}
The absence of $H_r < 5$ bodies in the cold classical belt strongly suggests that pebble accretion did not play a significant role in the formation of these objects. Compared to the primordial belt, which has an estimated mass of ${\sim}20\,m_\oplus$, the cold belt is four orders of magnitude less massive \citep{Kavelaars.2020}.  If accretion is local, as in the turbulent model, the far smaller number of bodies in the cold belt implies a significantly reduced pebble density ($\rho_\mathrm{peb}$), which would make growth by pebble accretion ineffective. Alternatively, in the laminar scenario the number of pebbles that could have drifted through would be much greater. However, the drawbacks outlined in the previous section would only intensify. The greater distance of the cold belt (${\sim}$40 au instead of ${\sim}$20 au) and the correspondingly higher aspect ratio $h$, would increase the characteristic growth timescale $t_0/\tau_s$, the pebble drift timescale $t_\mathrm{drift}$ (\eq{tdrift}) and the total mass of the pebble reservoir ($m_\mathrm{needed}$; \eq{m-needed}) to levels that become unrealistic. In addition, the conditions in the cold belt may have been such to inherently prevent pebble accretion from being triggered, i.e., $m\ll m_\ast$. The brightest cold classical is the binary (79360) Sila–Nunam \citep{Huang.202269} with an estimated system mass of $1.1 \times 10^{22}$~g \citep{Grundy.2012} ($D\approx280\,\mathrm{km}$). From \fg{massgain}, it is evident that these bodies barely fulfill the criterion for pebble accretion. While the specific conditions in the cold classical belt may differ, the parameters involved in setting the value of $m_\ast$ do not help. First, the higher $\eta$ would increase $m_\ast$ and, second, a lower $\tau_s$, while it would bring down $m_\ast$, would slow down pebble accretion to levels that make it unviable. In conclusion, it is highly improbable that the bodies in the cold belt grew substantially from a passing swarm of pebbles.

\subsection{Formation time constraints}
Recently, \citet{CanasEtal2024} also investigated the growth of TNOs by pebble accretion. Their approach is complementary to ours. In particular, their study emphasizes the composition and internal density of TNOs. A key constraint is that small TNOs of radius $50{-}100$ km could not have formed too quickly as radiogenic heating (primarily Al-26 decay at a half-life of 0.7 Myr \citealt{NorrisEtal1983}) would have melted these bodies and reduced their porosity. This outcome is in contrast with observationally inferred densities of TNOS, implying that these TNOs have retained significant porosity \citep{BiersonNimmo2019}.

\citet{CanasEtal2024} addressed this conundrum by hypothesizing that the initial distribution of TNOs formed through streaming instabilities dominated by large, ice-rich pebbles. Processes such as photodesorption could have resulted in a compositional dichotomy where ice is effectively transferred to the larger particles that settled to the midplane layers that were shielded from stellar UV irradiation. \citet{CanasEtal2024} argue that streaming instability would favor the larger ice-rich pebbles over the smaller Al-26 rich ones. Hence, the first-generation planetesimals were low in Al-26 and the early heating problem outlined by \citet{BiersonNimmo2019} is avoided. Subsequently, these bodies accreted smaller, more water-poor, and denser pebbles over much longer timescales when heating by Al-26 decay is less significant.

Alternatively, TNOs may have formed late, only after several half-lives of Al-26 \citep{BiersonNimmo2019}. This scenario implies that the conditions necessary for the primordial belt to form only emerged after several million years. Subsequently, pebble accretion could proceed without significant heating constraints. However, it would then be limited by the remaining time of the gaseous disk. Formation in a ring setting, where accretion timescales are short ($\sim$1 Myr) is then the preferred scenario.

\subsection{TNOs formation by planetesimal accretion}
An alternative to formation of TNOs by pebbles is growth among the planetesimal population itself \citep[e.g.,][]{KenyonBromley2004}. In this scenario accretion first proceeds through a rapid runaway growth phase \citep{WetherillStewart1989}, followed by a slower oligarchic growth phase \citep{KokuboIda1998}.
Runaway growth conditions require that the initial population of bodies of mass $m_0$ is dynamically quiescent (zero eccentricity), such that gravitational focusing becomes more effective for higher mass bodies. During runaway growth, a small number of bodies of mass much higher than $m_0$ emerge on a timescale that is a fraction of the collision timescale among the initial population, $t_\mathrm{col,0}$. The distribution evolves into a mass spectrum $dN/dm\propto m^{-2.5}$ \citep{KokuboIda1996,BarnesEtal2009,OrmelEtal2010i,MorishimaEtal2013}. During the subsequent oligarchic growth, the most massive bodies separate from the continuous mass distribution, resulting in a flattening of the cumulative mass function. However, these oligarchs dynamically excite the planetesimal population, increasing their eccentricities and significantly slowing down growth.

The similarity of the TNO mass distribution to a $p=-2.5$ power law (see \fg{cumulative}) has been interpreted as evidence for a runaway growth origin \citep{OrmelEtal2010,KenyonBromley2012,Morishima2017}. For the tapered-exponential distribution of \eq{Kavelaars2021}, it is natural to let $m_0$ coincide with the mass scale where most of the population mass resides (i.e., $m_0 \sim 4m_\mathrm{brk}\approx10^{21}\,\mathrm{g}$; see \app{TNOmass}), corresponding to a radius of about 50\, km. The initial collision timescale then evaluates
\begin{align}
	\label{eq:tcol0}
	t_\mathrm{col,0} & \sim \frac{\rho_\bullet R}{\Sigma_\mathrm{pltm}\Omega}
	= \frac{2\pi r W_r \rho_\bullet R}{(Gm_\odot/r^3)^{1/2}m_\mathrm{PB}}                                                                                                                                                                  \\ \nonumber
	                 & = 67\,\mathrm{Myr}\, \left( \frac{m_\mathrm{PB}}{20\,m_\oplus} \right)^{-1} \left( \frac{r}{20\,\mathrm{au}} \right)^{7/2} \frac{W_r/r}{0.2} \frac{\rho_\bullet}{1\,\mathrm{g\,cm^{-3}}} \frac{R}{50\,\mathrm{km}},
\end{align}
where $m_\mathrm{PB}$ is the mass of the primordial belt and $W_r/r$ its fractional width. Unlike pebble accretion, planetesimal-driven growth does not need to be completed before the gas disk disperses, but it must conclude before the onset of the dynamical instability. The survival of the binary Jupiter trojan Patroclus–Menoetius puts constraints on the level of dynamical interactions in the precursor primordial belt. Specifically, \citet{NesvornyEtal2018} find a surviving probability exponentially decaying by a factor of 10 for every 100 Myr. Studies simulating the timing of the giant planet instability are consistent with these constraints \citep{deSousaEtal2020}. Thus, while these constraints limit the time for collisional processing in the primordial belt, the runaway growth phase is likely to have been sufficiently rapid  (\eq{tcol0}) to establish the $p=-2.5$ mass spectrum.

However, the TNO size distribution flattens at high mass, suggesting that the high-mass bodies (Triton, Pluto, Eris) accreted significantly more mass than their low-mass counterparts. This characteristic is a hallmark of oligarchic growth, but it could also be achieved with pebble accretion (see, e.g., the \texttt{lami-1p-himass} model in \fg{sim-default}). Oligarchic growth, however, proceeds much more slowly than runaway growth and it may be challenging to meet the aforementioned timescale constraints. Alternatively, the runaway growth phase could have continued until Pluto-sized bodies formed, resulting in a continuous $p=-2.5$ mass spectrum, but with the subsequent stochastic implantation retaining a much higher fraction of TNO dwarf planets. In order to determine whether planetesimal or pebble accretion was the dominant formation mechanism, future simulations are needed to evaluate the scenarios under equivalent conditions. These efforts would greatly benefit from a more extensive and complete sample of TNOs.

\subsection{The definition of a dwarf planet}

The International Astronomical Union (IAU) defines a dwarf planet as ``an object in orbit around the Sun that is large enough to pull itself into a nearly round shape but has not been able to clear its orbit of debris.'' \footnote{\url{https://www.iau.org/public/themes/pluto/}}. While planetesimals that formed through SI can exhibit irregular shapes (as exemplified by Arrokoth's bi-lobed structure), significant pebble accretion would naturally lead to more spherical bodies through the slow settling of material and the isotropy of the accretion process. Therefore, a body grown by pebble accretion with $m_\mathrm{final} \gg m_\mathrm{init}$ is expected to have a nearly round shape. Specifically, if we set the mass gain factor at 10, $m_\text{final} = 10 m_\text{init}$, our simulations indicate that TNOs with masses greater than ${\sim}10^{-4}m_\oplus$ (or $6\times10^{23}$~g, comparable to Orcus) would fulfill the IAU requirement of dwarf planet (\fg{massgain}). Future stellar occultations of TNOs are essential to test this hypothesis.

\section{Conclusions}
\label{sec:conclude}
In this study, we investigated a scenario in which high-mass trans-Neptunian objects (TNOs) accreted pebbles before being implanted into their current locations. The characteristics of the TNO size distribution
suggest that the dynamically hot bodies featured processing at the high-mass end, where the distribution significantly flattens (see \fg{cumulative}). This finding aligns with the pebble accretion mechanism, which operates preferentially on the most massive bodies, those exceeding the critical mass threshold $m_\ast$ (\fg{timescales}). Focusing on a carefully selected sample of dynamically hot TNOs within 48 au, for which we argue the population is complete ($H_r<5$ for bodies in the Minor Planet Survey and $H_r<8.3$ for bodies in \ossp), we performed an MCMC analysis to best reproduce the observed present-day size distribution. This analysis allowed us to constrain the key parameters that describe the pebble accretion process, specifically the pebble aerodynamic size ($\tau_s$) and the pebble-planetesimal velocity ($\Delta v$).

Our main conclusions are the following:
\begin{enumerate}
	\item The TNO size distribution tightly constrains the combination of the TNO-pebble relative velocity $\Delta v$ and pebble aerodynamic size $\tau_s$ in the primordial belt (\Tb{outputtable} and \eq{tausig}). This corresponding threshold mass $m_\ast$, implies that bodies above ${\sim}10^{22}\,\mathrm{g}$ have enjoyed significant growth by pebble accretion (see \fg{massgain}). While simulations featuring $\tau_s$ smaller than ${\sim}10^{-3}$ would formally fit the data, they imply very long accretion times, which rival the lifetime of the disk. Bodies exceeding $m_\ast$ increase their mass several tenfold and become dwarf planets (see \fg{massgain}).
	\item The data can be fitted with either a laminar or a turbulent velocity field. There is at present no evidence of a second pebble size. This stems mainly from the lack of bodies in the high-mass tail of the TNO distribution, which render the fit insensitive to this part.
	\item TNOs likely did not form in a laminar disk. In such an environment, the radial pressure gradient would imply the particle aerodynamic size $\tau_s \lesssim 10^{-3}$. Such small pebbles couple too well to the gas for pebble accretion to be effective. It would have made the process too slow in comparison to disk lifetimes. In addition, pebble accretion in smooth disks would put enormous demands on the total mass of pebbles that must have passed through the primordial belt (\eq{m-needed}, most of which do not get accreted). These issues would be mitigated, however, in a cold disk environment.
	\item Most likely, pebble accretion proceeded locally, under conditions where pebbles are entrained by pressure and could not escape from the birth ring. In this case, $\eta=0$ and the relative velocity is due to turbulence. These conditions would resemble the observed continuum millimeter emission from continuum rings by ALMA.  The inferred ($\sigma,\tau_s$) relation from this work can be combined with the constraints that the extent of the rings provide to give a preferred pebble aerodynamic size of $\tau_s\approx10^{-2}$. Pebble accretion could have been complete in ${\sim}0.1{-}1$ Myr.
\end{enumerate}

Future surveys like CLASSY \citep{Fraser.2022} and especially the advent of the Vera C. Rubin Observatory (LSST, \citealt{Collaboration.2021}) will greatly expand the TNO inventory by many times over. This will allow us to further constrain the pebble accretion process and deepen our understanding of planet formation in greater detail.

\begin{acknowledgements}
	This work is supported by the National Natural Science Foundation of China under grant No. 12473065, 12233004, and 12250610189. Y.H. acknowledges funding support from Tsinghua University and National Observatory of Japan. The authors appreciate the referee for a thoughtful report, B.~Gladman and E.~Kokubo for useful discussions, and
	J.-M. Petit for providing the original \ossp data.
\end{acknowledgements}

\bibpunct{(}{)}{;}{a}{}{,} 
\bibliographystyle{aa}
\bibliography{tno-pebble} 

\begin{appendix}
\section{TNO masses}
\label{app:TNOmass}
\newcommand{\haumeanote}{Haumea family members excluded from the P48 sample}
\citet{KavelaarsEtal2021} modeled the magnitude distribution of the cold belt as
\begin{equation}
	\label{eq:Ncum-Hr}
	N(<H_r) = 10^{\alpha' (H_r-H_0)} \exp\left[ -10^{-\beta(H_r-H_B)} \right].
\end{equation}
where $H_r$ is the \oss magnitude, $\alpha'$ is the slope at low brightness objects, $\beta$ describes the strength of the exponential tapering, which kicks in at $H_r\sim H_B$, and $H_0$ is a normalization parameter. If we define a mass-magnitude relation as
\begin{equation}
	\label{eq:mass-mag1}
	m = m_B 10^{-3H_r/5}
\end{equation}
we can relate the parameters appearing in \eq{Ncum-Hr} to the cumulative distribution defined in \eq{Kavelaars2021}:
\begin{equation}
	\gamma = \frac{5}{3}\beta;\quad
	\alpha = \frac{5}{3}\alpha';\quad
	\frac{m_B}{m_\mathrm{brk}} = 10^{\beta H_B/\gamma};\quad
	N_\mathrm{pre} = 10^{-\alpha' H_0} \left( \frac{m_B}{m_\mathrm{bkr}} \right)^\alpha.
\end{equation}
With $\alpha'=0.4$ and $\beta=0.252$ \citep{KavelaarsEtal2021}, we obtain $\alpha=0.67$ and $\gamma=0.42$.
We adopt the mass-magnitude conversion of \citet{PetitEtal2023}, where in \eq{mass-mag1}
\begin{equation}
	\label{eq:mass-mag2}
	m_B = 9.685\times10^{23}\,\mathrm{g} \left( \frac{\rho_\bullet \nu^{-1.5}}{1\,\mathrm{g\,cm^{-3}}} \right)
\end{equation}
with $\nu$ the albedo.  For the cold belt, we take $\nu=0.15$ and $\rho_\bullet=1\,\mathrm{g\,cm^{-3}}$ \citep{PetitEtal2023}, which, with $\beta=0.252$ and $H_B=8.1$ \citep{KavelaarsEtal2021}, amounts to a breaking mass of $m_\mathrm{bkr} = 2.3\times10^{20}\,\mathrm{g}$ and a prefactor  for the cold belt $N_\mathrm{cold} = N_\mathrm{pre} = 1.9\times10^4$. For reference, the total mass, integrating \eq{Kavelaars2021}, amounts to
\begin{equation}
	m_\mathrm{tot}
	= \frac{\Gamma[(1-\alpha)/\gamma]}{\gamma} m_\mathrm{bkr} N_\mathrm{pre}
	= 2.81 m_\mathrm{bkr} N_\mathrm{pre},
	\label{eq:m-tot}
\end{equation}
which equals $2.0\times10^{-3}\,m_\oplus$ for the cold belt. In addition, the characteristic mass of the population, defined as the ratio of the second to first moment of the mass distribution, is $m_\ast = 2\Gamma[(2-\alpha)/\gamma] /\Gamma[(1-\alpha)/\gamma] = 3.97 m_\mathrm{bkr}$.

For the hot belt, \citet{PetitEtal2023} have shown that the shape of the magnitude distribution is the same as the cold belt with the number elevated by a factor of 2.2. This would nevertheless amount to a different shape for the mass distribution (\eq{Kavelaars2021}) since the albedos of the hot and cold belts differ, as is commonly assumed. Specifically, we take $\nu=0.08$ and $\rho_\bullet=1\,\mathrm{g\,cm^{-3}}$ for bodies in P48. However, at high mass it is known that the mass-magnitude relationship represented by \eqs{mass-mag1}{mass-mag2} breaks (see \fg{massKBO}). Several TNOs have mass estimates (those in P48 are listed in \Tb{tnos-mpc}) and we can use them to fit a mass--radius relationship for the high-mass end:
\begin{equation}
	\label{eq:mass-mag2}
	m = m_\mathrm{tr} 10^{-p_\mathrm{hm}(H_r-H_\mathrm{tr})}\qquad (m>m_\mathrm{tr}).
\end{equation}
Thus, at the point $(H_\mathrm{tr}, m_\mathrm{tr})$ the mass-magnitude relationship transitions from the low-mass expression (\eq{mass-mag1}) to \eq{mass-mag2}, valid at high mass. Performing the fit, we obtain $H_\mathrm{tr}=4.12$ and $p_\mathrm{hm}=0.3638$.

\begin{figure}[tb]
	\includegraphics[width=\columnwidth]{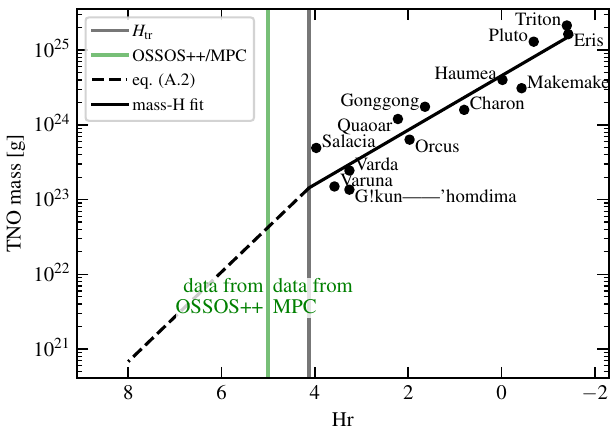}
	\caption{\label{fig:massKBO}Mass-magnitude relationship for the hot main belt. At low mass we adopt the relationship given by \citet{PetitEtal2023}, \eqs{mass-mag1}{mass-mag2} (dashed line). At high mass we fit a power-law relationship to the TNOs for which the mass is known.}
\end{figure}

\begin{table*}
	\centering
	\small
	\caption{\label{tab:tnos-mpc}All TNOs in the MPC sample.}
\begin{tabular}{llllllllll}
\hline
Name & $H_r$ & $a$ & mass & mass reference or Note & Name & $H_r$ & $a$ & Note \\
\hline
Triton               &    -1.40 &    30.07 & $2.14{\times}10^{25}$ & $^{1}$  & 1995.SM55            &     4.43 &    42.14 &  & $^{2}$ \\
Pluto                &    -0.69 &    39.40 & $1.30{\times}10^{25}$ & \citet{BrozovicJacobson2024}  & Chaos                &     4.43 &    46.07 &  &  \\
Makemake             &    -0.43 &    45.33 & $3.10{\times}10^{24}$ & \citet{ParkerEtal2018}  & 2012.VR128           &     4.46 &    41.64 &  &  \\
Haumea               &    -0.02 &    42.90 & $4.00{\times}10^{24}$ & \citet{RagozzineBrown2009}  & 2005.CB79            &     4.50 &    43.22 &  & $^{2}$ \\
Charon               &     0.80 &    39.40 & $1.59{\times}10^{24}$ & \citet{BrozovicJacobson2024}  & 2014.OE394           &     4.50 &    46.32 &  &  \\
Orcus                &     1.97 &    39.24 & $6.35{\times}10^{23}$ & \citet{GrundyEtal2019}  & 2002.KX14            &     4.51 &    38.64 &  &  \\
Quaoar               &     2.22 &    43.19 & $1.20{\times}10^{24}$ & \citet{MorgadoEtal2023}  & 2015.BZ518           &     4.52 &    46.71 &  &  \\
2002.AW197           &     3.25 &    47.13 &  &   & 2014.BV64            &     4.52 &    45.71 &  &  \\
Varda                &     3.26 &    45.62 & $2.45{\times}10^{23}$ & \citet{SouamiEtal2020}  & 2002.XV93            &     4.58 &    39.51 &  &  \\
Ixion                &     3.27 &    39.41 &  &   & 2014.TZ85            &     4.59 &    43.83 &  &  \\
2002.TX300           &     3.31 &    43.58 &  & $^{2}$  & 2010.VK201           &     4.59 &    43.50 &  & $^{2}$ \\
2002.MS4             &     3.42 &    41.66 &  &   & Huya                 &     4.60 &    39.20 &  &  \\
2005.RN43            &     3.49 &    41.68 &  &   & 1996.TO66            &     4.61 &    43.56 &  & $^{2}$ \\
2003.AZ84            &     3.55 &    39.53 &  &   & 2007.JH43            &     4.62 &    39.27 &  &  \\
Varuna               &     3.58 &    43.08 & $1.50{\times}10^{23}$ & \citet{LacerdaJewitt2007}  & 2008.AP129           &     4.64 &    41.81 &  & $^{2}$ \\
2002.UX25            &     3.65 &    43.02 &  &   & 2014.US224           &     4.64 &    47.14 &  &  \\
2005.UQ513           &     3.72 &    43.62 &  & $^{2}$  & 2015.AJ281           &     4.67 &    43.15 &  & $^{2}$ \\
2004.GV9             &     3.77 &    41.74 &  &   & 2014.FT71            &     4.69 &    43.34 &  &  \\
2003.VS2             &     3.79 &    39.70 &  &   & 2002.XW93            &     4.70 &    37.67 &  &  \\
2003.OP32            &     3.80 &    43.30 &  & $^{2}$  & 2014.UM33            &     4.71 &    43.35 &  &  \\
2010.KZ39            &     3.82 &    45.01 &  &   & Lempo                &     4.74 &    39.80 &  &  \\
2005.RR43            &     3.94 &    43.56 &  & $^{2}$  & 2014.BZ57            &     4.75 &    42.79 &  &  \\
Salacia              &     3.97 &    42.24 & $4.92{\times}10^{23}$ & \citet{GrundyEtal2019}  & 2014.HZ199           &     4.76 &    42.92 &  & $^{2}$ \\
2004.XA192           &     4.02 &    47.44 &  &   & 2014.VU37            &     4.76 &    40.97 &  &  \\
2013.FZ27            &     4.06 &    47.93 &  &   & 2014.JR80            &     4.77 &    39.39 &  &  \\
2003.UZ413           &     4.09 &    39.48 &  &   & 2011.OA60            &     4.77 &    40.80 &  &  \\
2004.TY364           &     4.13 &    39.15 &  &   & 2013.JW63            &     4.80 &    45.42 &  &  \\
2017.OF69            &     4.14 &    39.47 &  &   & 2014.XY40            &     4.84 &    47.53 &  &  \\
2004.UX10            &     4.18 &    39.29 &  &   & 2014.JP80            &     4.85 &    39.45 &  &  \\
2004.NT33            &     4.22 &    43.56 &  &   & 2014.WH509           &     4.91 &    44.30 &  &  \\
2014.WP509           &     4.24 &    45.23 &  &   & 2010.RO64            &     4.92 &    47.26 &  &  \\
2007.JJ43            &     4.29 &    47.72 &  &   & 2014.QW441           &     4.92 &    44.59 &  & $^{2}$ \\
2004.PF115           &     4.29 &    39.09 &  &   & 2003.QW90            &     4.93 &    44.09 &  &  \\
2013.XC26            &     4.31 &    42.41 &  &   & 2014.XR40            &     4.93 &    42.79 &  &  \\
2014.YA50            &     4.33 &    46.28 &  &   & 2015.AN281           &     4.96 &    41.75 &  &  \\
2010.FX86            &     4.35 &    46.54 &  &   & 2012.VB116           &     4.97 &    47.35 &  &  \\
2010.OO127           &     4.40 &    42.31 &  &   & 2005.CA79            &     4.98 &    47.84 &  &  \\
2009.YE7             &     4.40 &    44.62 &  & $^{2}$  & 2014.QA442           &     4.98 &    43.50 &  &  \\
2007.XV50            &     4.42 &    46.26 &  &   & 2014.AM55            &     4.98 &    47.15 &  &  \\
\hline
\end{tabular}
\tablefoot{$^{1}$\url{ https://ssd.jpl.nasa.gov/sats/phys_par/}; $^{2}$\haumeanote \ \citep{Proudfoot.2019}.}
\end{table*}

\end{appendix}

\end{document}